\documentclass[aps,12pt,final,notitlepage,onecolumn]{revtex4}

\usepackage{epsfig}
\usepackage{amsmath}
\usepackage{hyperref}

\begin{document}

\title{Photoproduction of $J/\psi$ and $\psi(2S)$
in proton--proton ultraperipheral collisions at the LHC}

\author{\firstname{Vadim} \surname{Guzey}}
\email{vguzey@pnpi.spb.ru}
\affiliation{Petersburg Nuclear Physics Institute (PNPI), \\
National Research Center ``Kurchatov Institute'', Gatchina, 188300, Russia}

\author{\firstname{Michael} \surname{Zhalov}}
\email{zhalov@pnpi.spb.ru}
\affiliation{Petersburg Nuclear Physics Institute (PNPI), \\
National Research Center ``Kurchatov Institute'', Gatchina, 188300, Russia}

\begin{abstract} 

Using the framework of leading order perturbative QCD along with 
the CTEQ6L1 gluon distribution of the proton and high-energy HERA data on the
$\gamma p \to J/\psi p$ and $\gamma p \to \psi(2S) p$ cross sections along with the 
corresponding H1 fit,
we obtain a good description of the rapidity dependence
of the cross sections of photoproduction of $J/\psi$ or $\psi(2S)$ mesons in proton--proton 
ultraperipheral collisions (UPCs) measured by the 
LHCb collaboration at the LHC. 
Within the same framework we also make predictions for the $pp \to pp J/\psi$ and  $pp \to pp \psi(2S)$ 
UPC cross sections at $\sqrt{s_{NN}}=8$ and 14 TeV.
We show that the possible contribution of the $p \to \Delta \gamma$ transition to the photon flux 
discernibly increases the $pp \to ppV$ UPC cross section and thus can affect the 
theoretical interpretation of results.

\end{abstract}

\maketitle

Ultraperipheral collisions (UPCs) of protons (ions) correspond to the situation when
they pass each other at the impact parameter $|\vec{b}|$ larger than the sum of
the hadron radii so that the strong interaction between the hadrons is suppressed
and they interact electromagnetically via emission of quasi-real photons.
Thus, UPCs at the LHC give a possibility to study photon--proton, photon--nucleus and 
photon--photon reactions at unprecedentedly high energies~\cite{Baltz:2007kq}.

Recently the LHCb collaboration at the LHC published results of an updated analysis of
exclusive photoproduction of $J/\psi$ and $\psi(2S)$ vector mesons in proton--proton
UPCs~\cite{Aaij:2014iea} at $\sqrt{s_{NN}}=7$ TeV, which supersedes and improves on 
the first measurement of this process made by LHCb~\cite{Aaij:2013jxj}.
The new data has much smaller experimental errors, which are of the order of 5\% for
$J/\psi$ and of the order of $15-20$\% for $\psi(2S)$, 
which allows one to better constrain and to some degree distinguish among various theoretical
approaches, see the discussion in~\cite{Aaij:2014iea}. The improved accuracy of the data also
requires a more accurate theoretical treatment, which should include effects that may contribute at the 
level of approximately 10\%. 
One such effect is the possibility for each proton to transform into a $\Delta$ 
while emitting a quasi-real photon in the $p \to \Delta \gamma$ transition~\cite{Baur:1998ay}.
Since the final proton and products of the $\Delta$ decay travel essentially along the beam pipe and, hence, are not detected
by the LHCb detector,
both $p \to \Delta \gamma$ and $p \to p \gamma$ transitions contribute to the photon flux
entering the calculation of the $pp \to ppV$ UPC cross section.

The aim of this note is twofold. First, we show that the leading order perturbative QCD provides the good 
description of the rapidity dependence and---to some extent---of the normalization of the 
$pp \to ppV$ cross section, where $V$ stand for $J/\psi$ or $\psi(2S)$, measured by the LHCb 
collaboration~\cite{Aaij:2014iea}.
 Second, we show how the additional
$p \to \Delta \gamma$ contribution to the photon flux affects the predicted $pp \to ppV$ cross section.

The cross section of photoproduction of $J/\psi$ and $\psi(2S)$ vector mesons in proton--proton
UPCs reads: 
\begin{eqnarray}
 \frac{d \sigma_{pp\to pp V}(y)}{dy} 
=r_{+}N_{\gamma/p}(y) (1+\delta(y))\sigma_{\gamma p\to V p}(y)+
r_{-}N_{\gamma/p}(-y)(1+\delta(-y))\sigma_{\gamma p\to Vp}(-y) \,,
\label{csupc}
\end{eqnarray}
where $V$ stands for $J/\psi$ or $\psi(2S)$;
$N_{\gamma/p}(y)=\omega dN_{\gamma/p}(\omega)/d\omega$ is the photon flux of the proton; 
$y = \ln(2\omega/M_V)$ is the vector meson rapidity, where $\omega$ is the photon 
energy in the reaction (laboratory) reference frame; 
$\sigma_{\gamma p\to V p}$ is the cross section of $J/\psi$ $[\psi(2S)$] photoproduction 
on the proton;
 $r_{+}$ and $r_{-}$ are absorptive corrections given by the rapidity gap survival
probabilities; $\delta(y)$ takes into account a possible additional contribution to the photon 
flux due to the $p \to \Delta$ transition. Below we consider each ingredient of Eq.~(\ref{csupc}) 
in detail.

The expression for the photon flux of a fast moving proton (ion) 
is well known from quantum electrodynamics, see, e.g.~\cite{Bertulani:1987tz}.
In practical applications, one often uses approximate expressions reproducing the exact result with 
a few percent accuracy. In this note, 
we use
the photon flux produced by a point-like (PL) particle passing a target at a minimal
impact parameter $b_{\rm min}$, which can be calculated analytically with the result:
\begin{equation}
N_{\gamma/p}(\omega)=\frac{2 \alpha_{\rm e.m.}}{\pi} 
\left[Y K_0(Y) K_1(Y)-\frac{Y^2}{2} \left(K_1^2(Y)-K_0^2(Y) \right) \right] \,,
\label{eq:flux_PL}
\end{equation}
where $\alpha_{\rm e.m.}$ is the fine-structure constant;
$K_0$ and $K_1$ are modified Bessel functions; 
$Y=\omega/\gamma_L b_{\rm min}$,
where $\gamma_L$ is the proton Lorentz factor and  
$b_{\rm min}=0.7$ fm~\cite{Nystrand:2004vn}.

In the leading logarithmic approximation of 
perturbative QCD, the high-energy $\gamma p \to V  p$
cross section is proportional to the gluon density of the proton 
$G_p(x,\mu^2)$ squared~\cite{Ryskin:1992ui}:
\begin{equation}
\frac{d \sigma_{\gamma p \to V p}}{dt}(W_{\gamma p},t=0)=C(\mu^2) [\alpha_s(\mu^2) x G_p(x,\mu^2)]^2
 \,,
\label{eq:cs}
\end{equation}
where $W_{\gamma p}$ is the invariant photon--proton center of mass energy; 
$\mu$ is the factorization scale;
 $x=M_V^2/W_{\gamma p}^2$ is the light-cone momentum fraction 
associated with the two-gluon exchange; 
$C(\mu^2)$ determines the cross section normalization, which depends 
on the wave function of the final charmonium state and approximations used in the calculation of the 
strong $\gamma p \to V  p$ amplitude.

In our approach, we treat Eq.~(\ref{eq:cs}) phenomenologically and determine the values of $\mu^2$ and
$C(\mu^2)$ from comparison to the high-energy HERA data on $J/\psi$~\cite{H1_2000,ZEUS_2002,H1_2005,H1_2013} 
and $\psi(2S)$~\cite{Adloff:2002re,Zulkaply:2012oaa} photoproduction on the proton.

In particular, in the $J/\psi$ case, we find~\cite{Guzey:2013qza} that Eq.~(\ref{eq:cs}) 
with $\mu^2 \approx 3$ GeV$^2$ and $C(\mu^2)=F^2(\mu^2)(1+\eta^2) R_g^2 \pi^3 \Gamma_{ee} M_{J/\psi}^3/(48 \alpha_{\rm em} \mu^8)$, where $F^2(\mu^2=3 \ {\rm GeV}^2)=0.48$ for the CTEQ6L1 leading-order gluon density of the 
proton~\cite{Pumplin:2002vw},
provides the good description of the $W_{\gamma p}$ dependence of the $\gamma p \to J/\psi p$ cross section and 
its normalization at $W_{\gamma p}=100$ GeV.
In the expression for $C(\mu^2)$~\cite{Ryskin:1995hz,Martin:2007sb}, 
$\Gamma_{ee}$ is the $J/\psi \to e^{+} e^{-}$ decay width;
$\eta$ is the ratio of the real to the imaginary parts 
of the $\gamma  p \to J/\psi p$ scattering amplitude and $R_g$ is the skewness factor taking into account 
the off-forward nature of this amplitude.
The factors of $\eta$ and $R_g$ are calculated from the $x$ dependence of $x G_p(x,\mu^2)$ at small $x$, 
see details in~\cite{Guzey:2013qza}.

To convert the differential cross section of Eq.~(\ref{eq:cs}) to the $t$ integrated
cross section, we used the conventional exponential parameterization of
the $t$ dependence of the $\gamma  p \to J/\psi p$ cross section:
\begin{equation}
\sigma_{\gamma p \to J/\psi p}(W_{\gamma p})=\frac{1}{B_{J/\psi}(W_{\gamma p})} \frac{d \sigma_{\gamma p \to J/\psi p}}{dt}(W_{\gamma p},t=0) \,,
\label{eq:cs_tint}
\end{equation}
where $B_{J/\psi}=4.5 +0.4 \ln (W_{\gamma p}/90 \ {\rm GeV})$, which is consistent with the HERA results~\cite{H1_2000,ZEUS_2002,H1_2005,H1_2013}.

In the $\psi(2S)$ case, the measured $W_{\gamma p}$ dependence of the diffractive $\gamma p \to \psi(2S) p$
cross section is found to be similar or possibly somewhat steeper than that for the $J/\psi$ 
cross section~\cite{Adloff:2002re,Zulkaply:2012oaa}.
Since the proton gluon density has an increasingly steeper $x$ dependence at small $x$ as
one increases the factorization scale $\mu$, the $W_{\gamma p}$ dependence of $\gamma p \to \psi(2S) p$
can be accommodated by Eq.~(\ref{eq:cs}) with $\mu^2 \approx 4$ GeV$^2$~\cite{Guzey:2014kka}.
In addition, 
using the result of the H1 analysis~\cite{Adloff:2002re} on the ratio of the $\psi(2S)$ and $J/\psi$ 
photoproduction cross sections on the proton, 
$\sigma_{\gamma p\to \psi(2S) p}/\sigma_{\gamma p\to J/\psi p}=0.166 \pm 0.007 ({\rm stat.}) \pm 0.008 ({\rm sys.}) \pm 0.007 ({\rm BR})$ on the interval $40 < W_{\gamma p} < 150$ GeV,
we fix the normalization of the $\sigma_{\gamma p \to \psi(2S) p}$ cross section as:
\begin{equation}
\sigma_{\gamma p \to \psi(2S) p}(W_{\gamma p}=100 \ {\rm GeV})=0.166\, \sigma_{\gamma p \to J/\psi p}(W_{\gamma p}=100 \ {\rm GeV}) \,,
\label{eq:cs_2S}
\end{equation}
where $\sigma_{\gamma p \to J/\psi p}(W_{\gamma p})$ is calculated using Eqs.~(\ref{eq:cs}) and (\ref{eq:cs_tint}).
To sum up, the $W_{\gamma p}$ dependence of the $\sigma_{\gamma p \to \psi(2S) p}(W_{\gamma p})$ cross section is 
calculated using Eq.~(\ref{eq:cs}) evaluated at $\mu^2 \approx 4$ GeV$^2$ 
and its normalization is fixed by Eq.~(\ref{eq:cs_2S}).

For the absorptive corrections $r_{+}$ and $r_{-}$ for $J/\psi$ and $\psi(2S)$
photoproduction in proton--proton UPCs, we used the results of~\cite{Jones:2013pga} and \cite{Jones:2013eda}, 
respectively. On average, $r_{\pm} \approx 0.8$.

Equation~(\ref{csupc}) also takes into account the possibility for each proton to transform into a $\Delta$ 
while emitting a quasi-real photon in the 
the $p \to \Delta \gamma$ transition.
The photon flux associated with the $p \to \Delta \gamma$ transition is~\cite{Baur:1998ay}:
\begin{eqnarray}
N_{\gamma/p \to \Delta}(\omega) &=& \frac{\alpha_{\rm e.m.}}{ 4 \pi} \frac{\mu^{\ast 2}}{9  m_p^2}
\left(\frac{M+m_p}{2 m_p} \right)^2 \Bigg[t_{\rm min} \Big\{\ln \frac{t_{\rm min}}{\Lambda^2+t_{\rm min}}
+\frac{11}{6}-\frac{2 t_{\rm min}}{\Lambda^2+t_{\rm min}} \nonumber\\
&+&\frac{3 t_{\rm min}^2}{2 (\Lambda^2+t_{\rm min})^2}
-\frac{t_{\rm min}^3}{3 (\Lambda^2+t_{\rm min})^3} \Big\}+\frac{\Lambda^8}{3 (\Lambda^2+t_{\rm min})^3} \Bigg]  \,,
\label{eq:Delta}
\end{eqnarray}
where $m_p$ and $M$ are proton and $\Delta$ masses, respectively; $\Lambda^2=0.71$ GeV$^2$;
$\mu^{\ast}=9.42$; $t_{\rm min}$ is the minimal momentum transfer, 
$t_{\rm min}=(\omega/\gamma_L)^2+(M-m_p)^2/\gamma_L^2+2 \omega (M-m_p)/\gamma_L$.

In Eq.~(\ref{csupc}), $\delta$ denotes the ratio of the photon fluxes due to the $p \to \Delta \gamma$ 
and $p \to p \gamma$ transitions:
\begin{equation}
\delta(y) \equiv \frac{N_{\gamma/p \to \Delta}(\omega)}{N_{\gamma/p}(\omega)} \,.
\label{eq:Delta2}
\end{equation}
In the LHC kinematics, the values of $\delta$ range from a few percent at small $\omega$ to 
$\delta \approx 0.1$ for $\omega={\cal O}(10 \ {\rm GeV})$ and to $\delta \approx 0.2$ for 
 $\omega={\cal O}(100 \ {\rm GeV})$.

\begin{figure}[t]
\centering
\epsfig{file=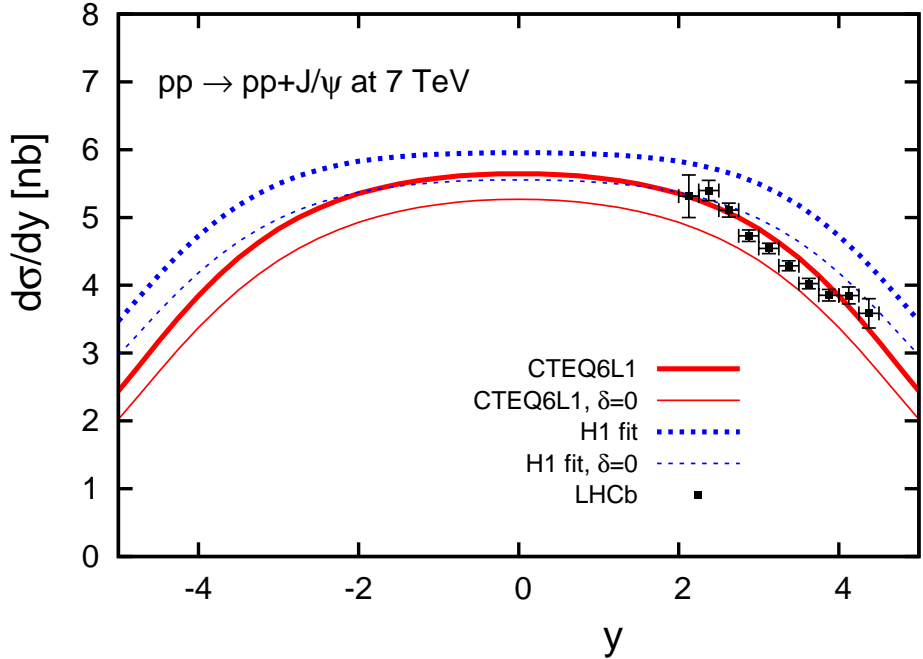,scale=1.7}
\caption{The cross section of $J/\psi$ photoproduction in proton--proton UPCs at $\sqrt{s_{NN}}=7$ TeV
as a function of the $J/\psi$ rapidity $y$. The theoretical predictions labeled by ``CTEQ6L1'' and ``H1 fit''
are compared to the LHCb data~\cite{Aaij:2014iea}.
}
\label{fig:Sigma_pp_2014}
\end{figure}

\begin{figure}[t]
\centering
\epsfig{file=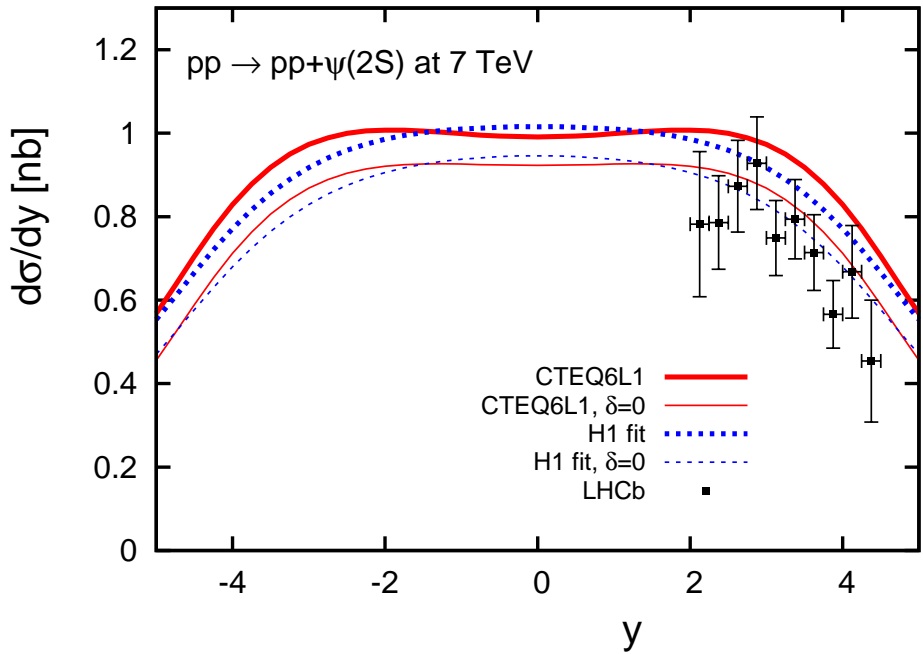,scale=1.7}
\caption{The cross section of $\psi(2S)$ photoproduction in proton--proton UPCs at $\sqrt{s_{NN}}=7$ TeV
as a function of the $\psi(2S)$ rapidity $y$. The theoretical predictions labeled by ``CTEQ6L1'' and ``H1 fit''
are compared to the LHCb data~\cite{Aaij:2014iea}.
}
\label{fig:Sigma_pp_2014_2s}
\end{figure}

Our results for $J/\psi$ and $\psi(2S)$ photoproduction in proton--proton UPCs at the LHC at 
$\sqrt{s_{NN}}=7$ TeV are shown and compared to the LHCb data~\cite{Aaij:2014iea}
in Figs.~\ref{fig:Sigma_pp_2014} and \ref{fig:Sigma_pp_2014_2s}, 
respectively. In the figures, we present $d \sigma_{pp\to pp V}(y)/dy$ of Eq.~(\ref{csupc}) 
as a function of the vector meson rapidity $y$. 
We show predictions corresponding to  $\sigma_{\gamma p \to V p}$ calculated using 
Eqs.~(\ref{eq:cs}), (\ref{eq:cs_tint}) and (\ref{eq:cs_2S}) with the CTEQ6L1 gluon density 
(curves labeled ``CTEQ6L1'')
and
also parameterized in the following simple form obtained in the H1 
analyses~\cite{Adloff:2002re,H1_2013} 
(curves labeled ``H1 fit''):
\begin{eqnarray}
\sigma_{\gamma p \to J/\psi p}^{\rm H1 \, fit}(W_{\gamma p})& =& N \left(\frac{W_{\gamma p}}{90 \ {\rm GeV}} \right)^{\lambda} \,, \nonumber\\
\sigma_{\gamma p \to \psi(2S) p}^{\rm H1 \, fit}(W_{\gamma p})& =& 0.166 N \left(\frac{W_{\gamma p}}{90 \ {\rm GeV}} \right)^{\lambda} \,,
\label{eq:H1fit}
\end{eqnarray}
where $N=81 \pm 3$ nb and $\lambda=0.67 \pm 0.03$.

As follows from the preceding, in our analysis we explored two viable possibilities 
for the $W_{\gamma p}$ dependence
of the $\sigma_{\gamma p \to \psi(2S) p}$ cross section: our leading order pQCD analysis corresponds approximately 
to
$\sigma_{\gamma p \to \psi(2S) p} \propto W_{\gamma p}^{0.9}$, while we used the result of the H1 fit corresponding
to $\sigma_{\gamma p \to \psi(2S) p} \propto W_{\gamma p}^{0.67}$.

In each case we consider two sets of theoretical curves: the thick upper-lying curves correspond to
the calculation including both $p \to p \gamma$ 
and $p \to \Delta \gamma$ transitions in the flux of equivalent photons 
and the thin lower-lying curves
correspond to the calculation where we set $\delta(y)=0$.

To obtain the experimental values for the $d \sigma_{pp\to pp V}(y)/dy$ cross section, we divided the 
published values of the differential cross section times the corresponding branching 
ratio~\cite{Aaij:2014iea} by the acceptance in each bin of $y$ and by the $V \to \mu^{+} \mu^{-}$
branching ratio. For the latter, we used ${\rm Br}(J/\psi \to \mu^{+} \mu^{-})=(5.93 \pm 0.06)$\% and
${\rm Br}(\psi(2S) \to \mu^{+} \mu^{-})=(7.7 \pm 0.8)\times 10^{-3}$~\cite{Nakamura:2010zzi}.
The experimental errors have been added in quadrature.

One can see from Fig.~\ref{fig:Sigma_pp_2014} that the theoretical description of the
$y$ dependence of the data is good. 
In addition, the leading order pQCD formalism employing the CTEQ6L1 gluon density also reproduces
 correctly the normalization
of the data in the $\delta(y) \neq 0$ case.
The H1 fit corresponding to the systematically larger $\sigma_{\gamma p \to J/\psi p}$ cross section 
overestimates the normalization of the $pp \to pp J/\psi$ cross section in the $\delta(y) \neq 0$ case but
agrees with the data much better in the $\delta(y) = 0$ case.

Turning to the $\psi(2S)$ case, 
one can see from Fig.~\ref{fig:Sigma_pp_2014_2s} that both the leading order pQCD framework and the H1 fit
reproduce the $y$ dependence of the $pp \to pp \psi(2S)$ cross section. 
As to the normalization, the calculation with $\delta(y)=0$ agrees with the data better than the
result of our calculation,
when we also include the $p \to \Delta \gamma$ transition.

Table~\ref{table:y_int} summarizes our predictions for the $pp \to pp J/\psi$ and $pp \to pp \psi(2S)$
cross sections integrated over the rapidity range $2 < y < 4.5$ taking into account the 
LHCb acceptance~\cite{Aaij:2014iea} and multiplied by the corresponding branching ratios for the two-muon 
decay~\cite{Nakamura:2010zzi}, $\sigma_{pp \to pp V \to pp\mu^{+} \mu^{-}}(2 < \eta_{\mu^{\pm}} < 4.5)$.

\begin{table}[h]
\begin{tabular}{|c|c|c|}
\hline
 & $J/\psi$ [pb] & $\psi(2S)$ [pb] \\
\hline
CTEQ6L1 & 298 & 7.9 \\
CTEQ6L1, $\delta(y)=0$ & 268 & 7.0 \\
H1 fit & 272 & 7.1 \\
H1 fit,  $\delta(y)=0$ & 311 & 6.7 \\
\hline
\end{tabular}
\caption{Predictions for the $pp \to pp J/\psi$ and $pp \to pp \psi(2S)$
cross sections integrated over the rapidity range $2 < y < 4.5$ taking into account the 
LHCb acceptance and multiplied by the the two-muon branching ratio, 
$\sigma_{pp \to pp V \to pp\mu^{+} \mu^{-}}(2 < \eta_{\mu^{\pm}} < 4.5)$.}
\label{table:y_int}
\end{table}

The values in Table~\ref{table:y_int} should be compared to LHCb result~\cite{Aaij:2014iea}:
\begin{eqnarray}
\sigma_{pp \to J/\psi \to \mu^{+} \mu^{-}}(2 < \eta_{\mu^{\pm}} < 4.5) & = & 291 \pm 7 \pm 19 \ {\rm pb} \,,
\nonumber\\
\sigma_{pp \to \psi(2S) \to \mu^{+} \mu^{-}}(2 < \eta_{\mu^{\pm}} < 4.5) & = & 6.5 \pm 0.9 \pm 0.4 \ {\rm pb} \,,
\end{eqnarray}
where the first uncertainty is statistical and the second one is systematic.
In the $J/\psi$ case, the best agreement between our predictions 
and the experimental value is found for the calculation using the CTEQ6L1 gluon distribution 
with $\delta(y) \neq 0$ and the H1 fit with $\delta(y) = 0$.
In the $\psi(2S)$ case, the best agreement between the experiment and theory is found in the case, when
the contribution of the $p \to \Delta \gamma$ transition is omitted ($\delta(y)=0$).

\begin{figure}[h]
\centering
\epsfig{file=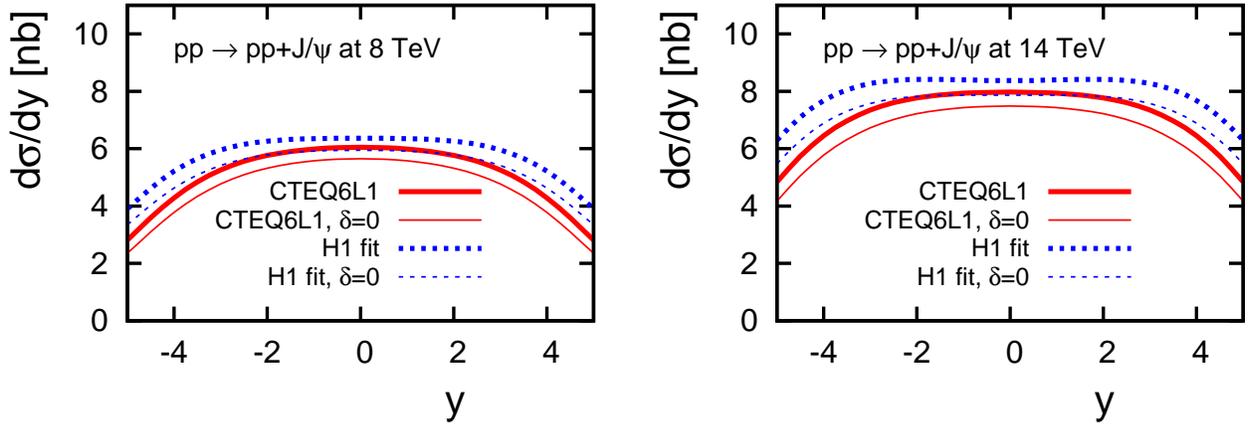,scale=1.7}
\caption{The cross section of $J/\psi$ photoproduction in proton--proton UPCs at $\sqrt{s_{NN}}=8$ TeV (left)
and $\sqrt{s_{NN}}=14$ TeV (right)
as a function of the $J/\psi$ rapidity $y$. Different curves correspond to different 
theoretical calculations explained in text.}
\label{fig:Sigma_pp_2014_14TeV}
\end{figure}

\begin{figure}[h]
\centering
\epsfig{file=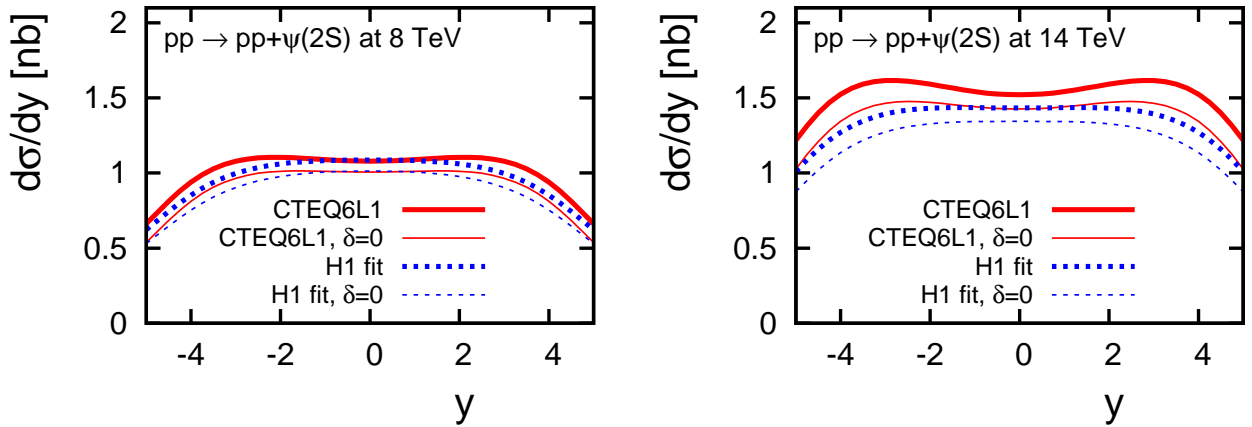,scale=1.7}
\caption{The cross section of $\psi(2S)$ photoproduction in proton--proton UPCs at $\sqrt{s_{NN}}=8$ TeV (left)
and $\sqrt{s_{NN}}=14$ TeV (right)
as a function of the $\psi(2S)$ rapidity $y$. 
Different curves correspond to different 
theoretical calculations explained in text.
}
\label{fig:Sigma_pp_2014_2s_14TeV}
\end{figure}

Figures~\ref{fig:Sigma_pp_2014_14TeV} and \ref{fig:Sigma_pp_2014_2s_14TeV} show our predictions for
the cross section of $J/\psi$ and $\psi(2S)$ photoproduction, respectively, in proton--proton UPCs at
$\sqrt{s_{NN}}=8$ TeV (left) and $\sqrt{s_{NN}}=14$ TeV (right). In these figures, different curves
correspond to different assumptions explained in text and used already in 
Figs.~\ref{fig:Sigma_pp_2014} and \ref{fig:Sigma_pp_2014_2s}.

Our analysis of high-energy exclusive photoproduction of $J/\psi$ and $\psi(2S)$ mesons on the 
proton, which is the underlying process in photoproduction of these mesons in proton--proton 
UPCs, is based on the two-gluon ladder exchange reaction mechanism in the leading logarithmic
approximation~\cite{Ryskin:1992ui} and, hence, is similar in the spirit to the analyses of 
Refs.~\cite{Jones:2013pga,Jones:2013eda}.
However, our approaches differ in implementation and details. In Ref.~\cite{Jones:2013pga},
combing the $\gamma p \to J/\psi p$ HERA data and the $pp \to pp J/\psi$ LHCb data, the leading and 
next-to-leading order gluon distributions in the proton have been determined. 
Using the obtained gluon distribution, predictions for $pp \to pp \psi(2S)$ were made 
in~\cite{Jones:2013eda}. In our approach, we use the leading order CTEQ6L1 gluon distribution and choose the 
factorization scale $\mu$ to reproduce the $W_{\gamma p}$ dependence of the $\gamma p \to J/\psi p$
and $\gamma p \to \psi(2S) p$ cross sections measured at HERA. As a result, we find that
$\mu^2 \approx 3$ GeV$^2$ for $J/\psi$ and $\mu^2 \approx 4$ GeV$^2$ for $\psi(2S)$, which are somewhat higher
than the leading-order values of $\mu^2=M_{J/\psi}^2/4=2.4$ GeV$^2$ for $J/\psi$
and $\mu^2=M_{\psi(2S)}^2/4=3.4$ GeV$^2$ for $\psi(2S)$ used in~\cite{Jones:2013pga,Jones:2013eda}.
While this difference in the $\mu^2$ values is not large, it affects the $y$ dependence of the
predicted $pp \to pp V$ UPC cross section.

Also, while we fix the normalization of the $\gamma p \to \psi(2S) p$ cross section using the HERA data,
which in turn determines the normalization of the $pp \to pp \psi(2S)$ cross section,
it is predicted in~\cite{Jones:2013eda} and, hence, can be used to assess the accuracy of the used framework
in the $\psi(2S)$ case.

Photoproduction of $J/\psi$ and $\psi(2S)$ in proton--proton UPCs was also considered in the
$k_t$-factorization approach~\cite{Schafer:2007mm,Cisek:2014ala}.
Using different models for the unintegrated gluon distribution, it was found~\cite{Cisek:2014ala} that the
rapidity dependence and normalization of the $pp \to pp J/\psi$ and $pp \to pp \psi(2S)$ cross sections measured
by the LHCb collaboration is best described by an unintegrated gluon distribution including nonlinear effects
and by the eikonal absorption factor of the order of $0.7$. Note that this corresponds to much stronger
absorption than that used in our analysis and in the analyses of Refs.~\cite{Jones:2013pga,Jones:2013eda}.
At the same time, since the simple H1 parameterization, see Eq.~(\ref{eq:H1fit}), leads to the similarly
good description of the LHCb data, a good agreement with the LHCb data within the $k_t$-factorization approach
precludes drawing a definite conclusion about an onset of saturation~\cite{Cisek:2014ala} .

Photoproduction of $J/\psi$ in proton--proton UPCs was also considered in the dipole 
approach~\cite{Goncalves:2007sa}. The resulting predictions are in a broad agreement with the LHCb data,
see the discussion in~\cite{Aaij:2014iea}.

In summary,
we showed that the framework of leading order perturbative QCD, where we used as an example 
the CTEQ6L1 gluon distribution of the proton, provides the good description of the rapidity dependence
of the cross sections of photoproduction of $J/\psi$ or $\psi(2S)$ mesons in proton--proton UPCs measured by the 
LHCb collaboration at the LHC. In addition, constraining the normalization of the $\gamma p \to J/\psi p$ and
 $\gamma p \to \psi(2S) p$ cross sections by the high-energy HERA data, allows us also to correctly reproduce
the normalization of the $pp \to ppV$ cross section. A similarly good description of the LHCb data
on the $pp \to ppV$ cross section
 is obtained using the H1 fit to the $\gamma p \to J/\psi p$ and
 $\gamma p \to \psi(2S) p$ cross sections. 
Using the same framework, we also made predictions for the  $pp \to pp J/\psi$ and  $pp \to pp \psi(2S)$ 
UPC cross sections for
$\sqrt{s_{NN}}=8$ and 14 TeV.
We showed that the contribution of the $p \to \Delta \gamma$ transition to the flux of equivalent photons
discernibly increases the $pp \to ppV$ UPC cross section and thus can affect its theoretical 
interpretation in the 
situation, when the final hadron is not detected as is the case for the LHCb detector.

\end{document}